\documentclass[conference]{IEEEtran}
\usepackage[utf8]{inputenc} 
\usepackage[T1]{fontenc}
\usepackage{url}
\usepackage{hyperref}
\usepackage{bm}
\usepackage{ifthen}
\usepackage{cite}

\usepackage{amssymb}
\usepackage[cmex10]{amsmath} 

\usepackage{graphicx}
\usepackage{color}

\def\t#1{\tilde{#1}}
\def\bb#1{{\mathbb #1}}
\def\r#1{{\rm #1}}

\IEEEsettopmargin{t}{0.8in}

\usepackage{comment}

\title{Deep Learning-Aided Projected Gradient Detector for Massive Overloaded MIMO Channels}
\author{
  \IEEEauthorblockN{Satoshi Takabe\IEEEauthorrefmark{1}\IEEEauthorrefmark{2}, 
 		Masayuki Imanishi\IEEEauthorrefmark{1}, 
                Tadashi Wadayama\IEEEauthorrefmark{1},
                and Kazunori Hayashi\IEEEauthorrefmark{3}}
  \IEEEauthorblockA{\IEEEauthorrefmark{1}%
		Nagoya Institute of Technology,
		Gokiso, Nagoya, Aichi 466-8555, Japan,\\
 		s\_takabe@nitech.ac.jp, 29414024@stn.nitech.ac.jp, wadayama@nitech.ac.jp} 
  \IEEEauthorblockA{\IEEEauthorrefmark{2}%
  		RIKEN Center for Advanced Intelligence Project,
  		Nihonbashi, Chuo-ku, Tokyo 103-0027, Japan
                }
  \IEEEauthorblockA{\IEEEauthorrefmark{3}%
		Graduate School of Engineering, Osaka City University,
                Sugimoto, Sumiyoshi-ku, Osaka, 558-8585, Japan,\\
                kazunori@eng.osaka-cu.ac.jp
                }

}

\begin{document}
%
\maketitle

\begin{abstract}
The paper presents a deep learning-aided iterative detection algorithm for massive overloaded MIMO systems.
Since the proposed algorithm is based on the projected gradient descent method with 
trainable parameters, it is named as  trainable projected descent-detector (TPG-detector).
The trainable internal parameters can be optimized with standard deep learning techniques 
such as back propagation and stochastic gradient descent 
algorithms. This approach referred to as data-driven tuning 
brings notable advantages of the proposed scheme such as fast convergence.
The numerical experiments show that TPG-detector achieves comparable detection performance
to those of the known algorithms for massive overloaded MIMO channels
with lower computation cost.
\end{abstract}


\section{Introduction}\label{sec_intro}

Multiple input multiple output (MIMO) systems have attracted great interests 
because they potentially achieve high spectral efficiency in wireless communications.
Recently, as a consequence of high growth of mobile data traffic, 
\emph{massive} MIMO is regarded as a key technology in the 5th generation (5G) wireless network standard~\cite{Yang}.
In massive MIMO systems, tens or hundreds of antennas are used in the transmitter and the receiver.
This fact complicates the detection problem for MIMO channels because the computational complexity of a MIMO detector, in general, 
increases as the numbers of antennas grow.  A practical massive MIMO 
detection algorithm should possess both low energy consumption and low computational complexity in addition to   
reasonable bit error rate (BER) performance.

In a down-link massive MIMO channel with mobile terminals, a transmitter in a base station can have many
antennas but a mobile terminal cannot have such a number of receive antennas because of the restrictions on 
cost, space limitation, and power consumption.
This scenario is known as the \emph{overloaded} (or \emph{underdetermined}) scenario.
Development of an overloaded MIMO detector with computational efficiency and reasonable BER performance 
is a highly challenging problem 
because conventional naive MIMO decoders such as 
the minimum mean square error (MMSE) detector~\cite{Shnidman} 
exhibit poor BER performance for overloaded MIMO channels,
and an optimal detection based on the exhaustive search is evidently computationally intractable. 

Several search-based detection algorithms such as slab-sphere decoding~\cite{SSD}
and enhanced reactive tabu search (ERTS)~\cite{ERTS} have been proposed for overloaded MIMO channels.
Though these schemes show excellent detection performance, 
they are computationally demanding, 
and it may prevent us from implementing them into a practical massive overloaded MIMO system.
As a computationally efficient approach based on the $\ell_1$-regularized minimization,
 Fadlallah et. al proposed a detector using a convex optimization solver~\cite{Fad1}.
Recently, Hayakawa and Hayashi~\cite{IW-SOAV2} proposed an iterative detection algorithm with practical computational complexity 
called  \emph{iterative weighted sum-of-absolute value} (IW-SOAV) optimization (see also ~\cite{IW-SOAV1}).
The algorithm is based on the SOAV optimization~\cite{SOAV} for sparse discrete signal recovery.
In addition, the algorithm includes a re-weighting process 
based on the log-likelihood ratio, which improves the detection performance.
The IW-SOAV provides the state-of-the-art BER performance among overloaded MIMO detection algorithms 
with low computational complexity.

The use of deep neural networks has spread to numerous fields such as image recognition~\cite{Image1}
with the progress of computational resources.
It also gives a great impact on design of algorithms for wireless communications and signal processing~\cite{Com1,Com2}.
Gregor and LeCun first proposed the learned iterative shrinkage-thresholding algorithm (LISTA)~\cite{LISTA}, which 
exhibits better recovery performance than that of the original ISTA~\cite{ISTA2} for sparse signal recovery problems.
Recently, the authors proposed the \emph{trainable ISTA} (TISTA)~\cite{TISTA}
yielding significantly faster convergence than ISTA and LISTA. 

TISTA includes several trainable internal parameters and these parameter are tuned with 
standard deep learning techniques such as back propagation and stochastic gradient descent (SGD) algorithms.
From our research work on TISTA \cite{TISTA} and several additional experiments 
(an example will be presented in Section~\ref{sec_tpg}), 
we encountered a phenomenon that 
the convergence to the minimum value 
 is accelerated with appropriate parameter embedding for several numerical optimization 
 algorithms such as the projected gradient descent method and the proximal gradient method.
We call this phenomenon {\em data-driven acceleration} of convergence.

Most of known acceleration techniques for gradient descent algorithms
such as the momentum methods
do not care about the statistical nature of the problems.
On the other hand, the data-driven acceleration is obtained 
by learning the statistical nature of the problem, i.e., 
stochastic variations on the landscape of the objective functions.
The internal parameters controlling the behavior of the algorithm 
are adjusted to match the typical objective function via training processes.
The data-driven acceleration is especially advantageous in implementation of 
detection algorithms because it reduces the number of
iterations without sacrificing the detection performance. This makes the algorithm
faster and more power efficient.

The goal of this paper is 
to propose a novel detection algorithm for massive overloaded MIMO systems,
which is called \emph{Trainable Projected Gradient-Detector} (TPG-detector).
The proposed algorithm is based on the projected gradient descent method with 
trainable parameters. We have confirmed that data-driven acceleration improves the 
detection performance and the convergence speed.
Though deep learning architectures for massive MIMO systems were recently proposed
as deep MIMO detectors (DMDs) in~\cite{DMD1,DMD2},
no deep learning-aided iterative detectors 
for massive \emph{overloaded} MIMO channels have been proposed as far as 
the authors are aware of. Furthermore, an application of data-driven tuning 
to MIMO detectors has not yet been studied in the related literatures.

\vspace{-0.2\baselineskip}
\section{Data-Driven Tuning}

We here introduce the concept 
of the {\em data-driven tuning of numerical optimization algorithms}, whose origin 
is the work by Gregor and LeCun~\cite{LISTA}.
For improving the performance of a numerical optimization algorithm, 
several trainable parameters can be embedded in the algorithm.
By unfolding an iterative process of a numerical optimization algorithm, we have
a multilayer signal-flow graph that is similar to a deep neural network.
Since each component of the signal-flow graph is differentiable,
these trainable parameters can be adjusted by standard deep learning techniques.
The training data can be randomly generated according to a 
channel model.

\begin{figure}[ht]
\begin{center}
\includegraphics[scale=0.38]{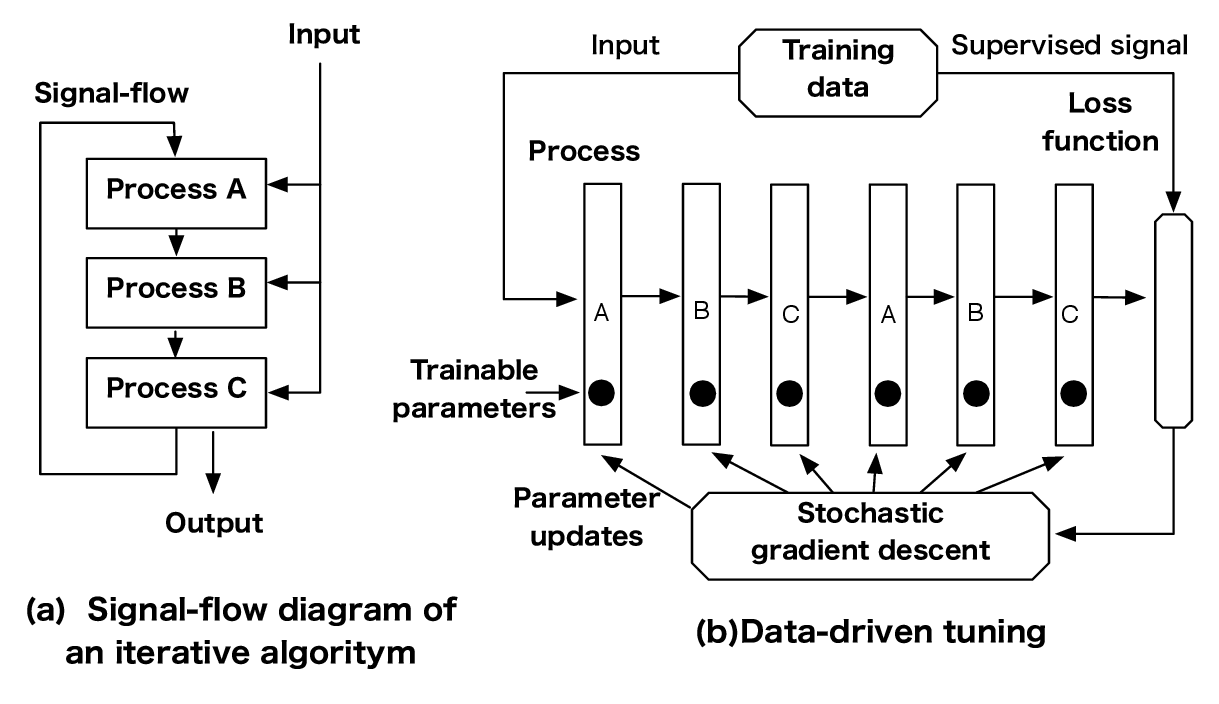}	
\caption{(a) A signal-flow diagram of an iterative algorithm, (b) Data-driving tuning based on an unfolded signal-flow graph with a loss function}
\label{signalflow}
\end{center}
\end{figure}

Figure \ref{signalflow} (a) illustrates a signal-flow diagram of
an iterative numerical optimization algorithm where Processes  A, B, and C are processes 
whose input/output relationships are expressed with differentiable functions.
By unfolding the signal-flow diagram, 
we obtain a signal-flow graph similar to a multilayer neural network (Fig. \ref{signalflow}(b)).

Each process contains trainable parameters that are represented by the black 
circles in Fig. \ref{signalflow}(b).  The trainable parameters can control
behavior of the processes A, B, and C.
Appending a 
loss function, e.g., the squared loss function,  at the end of the unfolded signal-flow graph, 
we are ready to feed randomly generated training data to the graph.
We can apply back propagation and 
a SGD type parameter update (SGD, RMSprop, Adam, etc.)
 to optimize the parameters.

\section{Data-Driven Acceleration}\label{sec_tpg}

We now discuss a simple example of the data-driven acceleration in more detail.
As a toy model closely related to the MIMO channel, we consider
a quadratic programming problem with binary variables.

Let us consider a simple quadratic optimization problem 
\begin{equation} 
	\mbox{minimize }\frac{1}{2}	\| A x - y \|_2^2 \mbox{ subject to } x \in \{-1, +1\}^n, \label{eq_QPB}
\end{equation}
where  $A\in\mathbb{R}^{n\times n}$ is a given matrix and $\| \cdot \|_2$ represents the Euclidean norm.
We assume that $y$ is stochastically generated as $y = A \tilde x + w \in \mathbb{R}^n$ where
$\tilde x$ is a vector randomly sampled from $\{-1, +1\}^n$ uniformly at random and
$w\in \mathbb{R}^n$ 
consists of i.i.d. Gaussian random variables with zero mean and variance $\sigma^2$.
The optimization problem is essentially same as the maximum likelihood estimation rule for the Gaussian 
linear vector channel.

Since solving this problem is known as an NP-hard problem in general, we need to solve 
the problem approximately.
We here exploit a valiant of the projected gradient (PG) algorithm 
to solve (\ref{eq_QPB}) approximately.
The PG algorithm can be described by the recursive formula:
\begin{eqnarray}
  \label{eq:pg_1}
  r_t &=& s_t + \gamma A^{T}(y - A s_t),\\
  \label{eq:pg_2}  
  s_{t+1} &=& \tanh\left(\alpha {r_t}\right),
\end{eqnarray}
where $t=1,\dots, T$ and $\tanh(\cdot)$ is calculated element-wisely.

The PG algorithm consists of two computational steps for each iteration.
In the gradient descent step~(\ref{eq:pg_1}), a search point 
moves to the opposite direction to {the gradient of the objective function, i.e.,
$\nabla \frac 1 2 \|A x - y\|_2^2 = - A^T(y - A x)$.}
The parameter $\gamma$ controls the step size causing critical influence on the convergence behavior.
In the projection step~(\ref{eq:pg_2}), {\em soft-projection} based on the 
hyperbolic tangent function is applied to the search point
to obtain a new search point nearly rounded to binary values.
Precisely speaking, the projection step  is not the projection to 
the binary symbols $\{-1, +1\}$. 
This is because the true projection 
to discrete values results in insufficient convergence behavior in a minimization process.
The parameter $\alpha$ controls the softness of the soft projection.
Note that this type of nonlinear projection has been commonly used 
in several iterative multiuser detection algorithms such as the soft parallel interference canceller~\cite{PIC}.

According to the data-driven tuning framework,
we can embed trainable parameters into  the PG algorithm.
The trainable PG (TPG) algorithm is based on the recursion 
\begin{eqnarray}
  \label{eq:tpg_1}
  r_t &=& s_t + \gamma_t A^{T}(y - A s_t),\\
  \label{eq:tpg_2}  
  s_{t+1} &=& \tanh\left(\alpha {r_t}\right).
\end{eqnarray}
The trainable parameters $\{\gamma_t \}_{t=1}^{T}$ play a key role in the gradient descent step 
by adjusting its step size adaptively.
For simplicity, the parameter $\alpha$ is fixed and treated as a hyper parameter.

As described in the last section, 
the parameters $\{\gamma_t\}_{t=1}^T$ are optimized by the standard mini-batch training.
The $i$th training data $d^{i} \triangleq (x^i, y^i)$ is generated randomly, 
i.e., $x^i \in \{-1, +1 \}^n$ is generated uniformly at random and
the corresponding $y^i$ is then generated according to $y^i = Ax^i + w$ with a given $A$.
A mini-batch consists of $D$ training data $\mathcal{D}\triangleq \{d^{1}, d^{2},\dots, d^{D}\}$.
In the following experiment,
a matrix $A$ is randomly generated for each mini-batch.
Each element of $A$ follows the Gaussian PDF with mean $0$ and variance $1$.

For each round of a training process, we feed these mini-batches 
to the TPG algorithm to minimize the squared loss function
$  L(\Theta_t) \triangleq D^{-1} \sum_{d^i \in \mathcal{D}} \| x^i -  \hat x^t(y^i) \|^{2}_{2}$, 
where $\hat x^t(y) \triangleq  s_{t+1}$ is the output of the TPG algorithm with $t$ iterations and
$\Theta_t\triangleq \{\gamma_1,\dots,\gamma_t\}$ is a set of trainable parameters up to the $t$th round.
A back propagation process evaluates the gradient $\nabla   L(\Theta_t)$ and 
it is used for updating the set of parameters as $\Theta_t:= \Theta_t + \Delta$ where 
$\Delta$ is determined by a SGD type algorithm such as the Adam optimizer. 

\begin{figure}[tb]
  \centering
  \includegraphics[width=7.3cm]{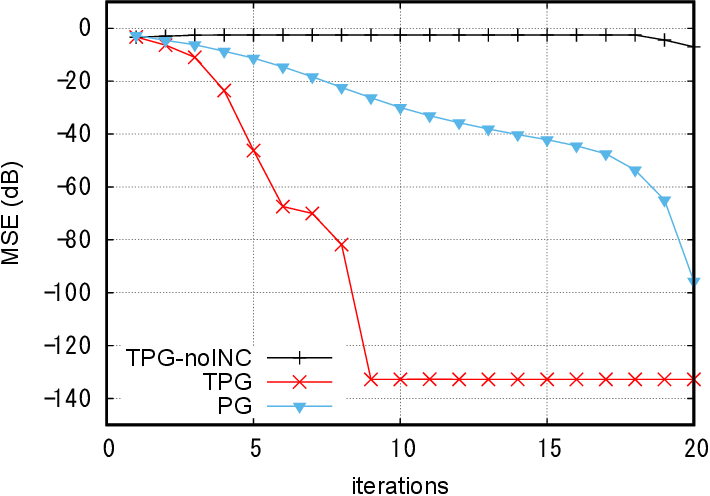}
  \caption{MSE as a function of iteration steps. The curve (PG) represents 
  the MSE of the plain PG algorithm with $\gamma=6.5\times 10^{-4}$.
The curve (TPG) corresponds to the MSE of the TPG algorithm and
{the curve (TPG-noINC) corresponds to the TPG algorithm without incremental training}.}
  \label{fig:tpg_iter}
\end{figure}

It should be remarked that a simple shingle-shot training for a whole process by letting $t = T$
does not work well (see also Fig.~\ref{fig:tpg_iter}) because the vanishing gradient phenomenon prevents appropriate parameter updates
due to the fact that the derivative of the soft projection function (\ref{eq:tpg_2}) becomes nearly zero almost everywhere.
In order to avoid the vanishing gradient phenomenon, we use an alternative approach, i.e., 
 \emph{incremental training} as TISTA~\cite{TISTA}.
In the incremental training, the parameters $\{\gamma_t\}_{t=1}^T$ are sequentially 
trained from $\Theta_1$ to $\Theta_T$ in an incremental manner.
The details of the incremental training is as follows. At first,  $\Theta_1$ is trained by minimizing $L(\Theta_1)$.
After finishing the training of $\Theta_1$, 
the values of trainable parameters in $\Theta_1$ are copied to 
the corresponding parameters in $\Theta_2$.
In other words, the results of the training for $\Theta_1$ are taken over to $\Theta_2$ as the initial values.
For each round of the incremental training which is called a {\em generation}, $K$ mini-batches are processed.

We show the numerical demonstration of the TPG algorithm.
In the experiment, the noise variance is fixed to $\sigma^2=4.0$.
The number of iterations of the TPG algorithm is $T = 20$.
In the training process, we use $K=100$ mini-batches per generation.
The mini-batch size is set to ${D}=200$ and  Adam optimizer~\cite{Adam} with learning rate $0.0005$ 
is used for the parameter updates.
The initial value of the trainable parameters are given by  $\gamma_t=1.0\times 10^{-4}$ ($t=1,\dots,T$).
The softness parameter is fixed to $\alpha = 8.5$ for the TPG algorithm.

\begin{figure}[tb]
  \centering
  \includegraphics[width=7.3cm]{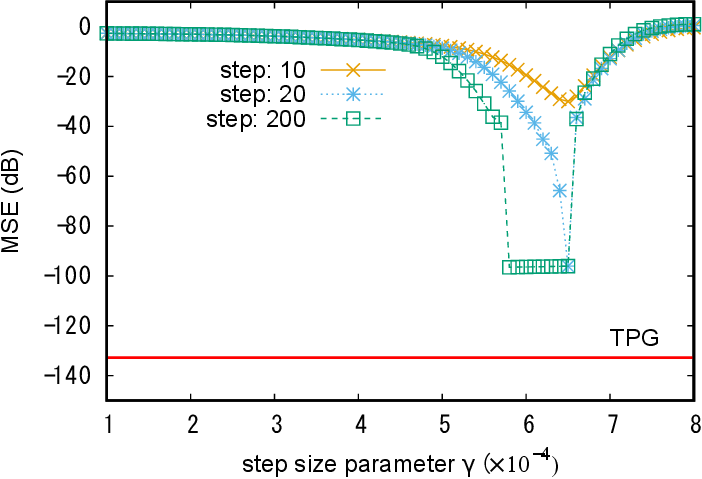}
  \caption{$\gamma$ dependence of the MSE performance in the plain PG algorithm.
  The horizontal solid line represents the MSE of TPG with $T=20$.}
  \label{fig:tpg_gamma}
\end{figure}

Figure~\ref{fig:tpg_iter} shows the mean squared error (MSE) as a function of iteration steps 
of the plain PG algorithm based on (\ref{eq:pg_1}), (\ref{eq:pg_2}) ($\alpha=6.0$, $\gamma=6.5\times 10^{-4}$) and
 the TPG algorithms based on (\ref{eq:tpg_1}), (\ref{eq:tpg_2}) (with/without incremental training).
The MSE is defined by $10 \log_{10} (E[||x - \hat x^t(y)||_2^2]/n)$ (dB)
and it is estimated from $10^4$ random samples of $A, y$ and $x$. 
The parameter $\gamma=6.5\times 10^{-4}$ in the plain PG algorithm is the optimal value 
for $T=20$ (See also Fig.\ref{fig:tpg_gamma}).
From Fig.~\ref{fig:tpg_iter}, we can observe that
the TPG algorithm provides much smaller MSEs than those of the plain PG algorithm.
The MSE of TPG achieves $-80$ dB at $t = 8$ but the PG yields the smaller MSE after $t = 19$.
Namely, TPG shows much faster convergence and 
it implies that the optimal parameter tuning drastically improves the convergence speed.
This is an example of the data-driven acceleration of convergence.
The effect of the incremental training can be confirmed by comparing 
the MSEs of the TPG algorithms with/without incremental training.
The MSE curve of {TPG-noINC} is almost flat, {indicating that the parameter tuning} is not successful
in the case of the TPG algorithm without incremental training.

In Fig.~\ref{fig:tpg_gamma}, we show the $\gamma$ dependence of the MSE performance in the plain PG algorithm.
We find that the parameter $\gamma$ must be selected carefully to obtain appropriate convergence.
In other words, the sweet spot of $\gamma$ is relatively narrow, i.e.,  close neighborhood of $6 \times 10^{-4}$
is only allowable choice for achieving  $-100$ dB at $T=200$. This means that 
optimization of the step size is critical even for the plain PG algorithm.
In addition, the TPG algorithm achieves the lower MSE performance (around $-130$ dB)
which cannot be achieved by the plain PG algorithm. This fact implies that having independent step size 
parameters for each iteration provides substantial improvement on the quality of the solution.

\section{Problem Setting for Overloaded MIMO Channels}\label{sec_set}

The section describes the channel model and introduces several definitions and notation.
The numbers of transmit and receive antennas are denoted by $n$ and $m$, respectively.
We only consider the overloaded MIMO scenario in this paper where $m <n$ holds.
It is also assumed that the transmitter does not use precoding and that the receiver perfectly knows 
the channel state information, i.e., the channel matrix.

Let $\t{x} \triangleq [\t{x}_1,\t{x}_2,\dots,\t{x}_n]^{T} \in \t{{\mathbb S}} ^{n}$ be 
a vector which consists of  a transmitted symbol $\t{x}_j$ ($j=1,\dots,n$) from the $j$th antenna.
The symbol $\t{\bb{S}} \subset \bb{C}$ represents a symbol alphabet, i.e., a signal constellation.
Similarly, $\t{y} \triangleq [\t{y}_1,\t{y}_2,\dots,\t{y}_m]^{T} \in \bb{C}^{m}$ denotes 
a vector composed of a received symbol $\t{y}_i$ $(i=1,\dots,m)$ by the $i$th antenna.
A flat Rayleigh fading channel is assumed here and the received symbols $\t{y}$ then reads
$
  \t{y} = \t{H}\t{x} + \t{w}, 
$
where $\t{w}\in \bb{C}^m$ consists of complex Gaussian random variables 
with zero mean and covariance $\sigma_{w}^2I$.
The matrix $\t{H} \in \bb{C}^{m \times n}$ is a channel matrix whose $(i,j)$ entry $\t{h}_{i,j}$ represents
a path gain from the $j$th transmit antenna to the $i$th receive antenna.
Each entry of $\t{H}$ independently follows the complex Gaussian distribution with zero mean and unit variance.
For the following discussion, it is convenient to derive an equivalent channel model defined over $\mathbb{R}$,
i.e.,  $y = Hx + w$, where 
    \begin{eqnarray} \nonumber
      y &\triangleq& \begin{bmatrix}
        \r{Re}(\t{y}) \\
        \r{Im}(\t{y})
        \end{bmatrix} \in \bb{R}^{M},\ 
      H \triangleq \begin{bmatrix} \label{eq:H_real}
        \r{Re}(\t{H}) & - \r{Im}(\t{H})\\
        \r{Im}(\t{H}) &  \r{Re}(\t{H})\\
      \end{bmatrix}, \\ \nonumber
      {x} &\triangleq& \begin{bmatrix}
        \r{Re}(\t{{x}}) \\
        \r{Im}(\t{{x}})
        \end{bmatrix} \in \bb{S}^{N}, \ 
      {w} \triangleq \begin{bmatrix}
        \r{ Re}(\t{{w}}) \\
        \r{ Im}(\t{{w}})
        \end{bmatrix}\in \bb{R}^{M},
    \end{eqnarray}
 and $(N,M) \triangleq (2n,2m)$.
The signal set $\bb{S}$ is the real counter part of $\t{\bb{S}}$.
The matrix $H \in \bb{R}^{M \times N}$ is converted from $\t{H}$.
 Similarly, the noise vector $w$ consists of i.i.d. random variables following the Gaussian distribution
 with zero mean and variance $\sigma^2_w/2$.
Signal-to-noise ratio (SNR) per receive antenna is then represented by
  $
  \mathrm{SNR}\triangleq E_s/N_0
  =  {N}/{\sigma_{w}^{2}}
  $,
  where $E_s\triangleq \bb{E}[||\t{H}\t{x}||^{2}_{2}]/m$ stands for the signal power per receive antenna and
  $N_0 \triangleq \sigma_w^2$ stands for the noise power per receive antenna.
Throughout the paper, we assume the QPSK modulation format, i.e., $\t{\bb{S}} \triangleq \{1+j,-1+j,-1-j,1-j\}$, 
which is equivalent to the BPSK modulation $\bb{S} \triangleq \{-1,+1\}$.

\section{Trainable Projected Gradient (TPG)-Detector}


The maximum likelihood estimation rule for the MIMO channel defined above 
is given by
\begin{equation} \label{mlprob}
	\hat x = \mbox{argmin}_{x \in \{-1, +1\}^N }	\| H x - y \|_2^2.
\end{equation}
This problem is a non-convex problem and finding the global minimum is computationally intractable for
a large scale problem.
Our proposal, TPG-detector, 
is based on 
the projected gradient method for solving the above non-convex problem approximately.
The process of TPG-detector is described by the following recursive formulas:
\begin{eqnarray}
  \label{eq:hoge_1}
  r_t &=& s_t + \gamma_t W(y - H s_t),\\
  \label{eq:hoge_2}  
  s_{t+1} &=& \tanh\left(\frac{r_t}{|\theta_t|}\right),
\end{eqnarray}
where $t ( = 1, \ldots, T)$ represents the index of an iterative step (or layer) and we set $s_1=0$ as the initial value.
This algorithm estimates a transmitted signal $x$ from the received signal $y$ and outputs 
the estimate $\hat x = s_{T+1}$ after $T$ iterative steps.

The steps (\ref{eq:hoge_1}) and (\ref{eq:hoge_2}) correspond to the gradient descent step and
to the projection step, respectively, as described in Section~\ref{sec_tpg}s.
The matrix $W$ in the gradient step (\ref{eq:hoge_1}) is the Moore-Penrose  
pseudo inverse matrix of $H$, i.e., $W \triangleq H^{T}(HH^{T})^{-1}$.
Precisely speaking, 
$W$ should be $H^T$ as in~(\ref{eq:tpg_1}).
However, we adopt the modification inspired by~\cite{OAMP} because 
this modification improves the BER performance of the proposed scheme.
As in the case in~(\ref{eq:tpg_2}), we use the hyperbolic tangent function as the soft projection.

The trainable parameters of TPG-detector  are $2T$ 
real scalar variables $\{\gamma_t\}_{t=1}^T$ and  $\{\theta_t\}_{t=1}^T$ 
in (\ref{eq:hoge_1}) and (\ref{eq:hoge_2}), respectively.
The parameters $\{\gamma_t\}_{t=1}^T$ in the gradient step
control the step size of a move of the search point.
In order to achieve fast convergence,
appropriate setting of these step size parameters is of critical importance as described in Section~\ref{sec_tpg}.
It should be remarked that similar constant trainable parameters 
are also introduced in the structure of TISTA~\cite{TISTA}.
The parameters $\{\theta_t\}_{t=1}^T$ {control} the softness of the soft projection 
in (\ref{eq:hoge_2}). 
One of the advantages of TPG-detector is that the number of trainable parameters is small, i.e., $O(T)$, 
and it leads to fast and stable training processes.
For example, the number of trainable parameters of TPG-detector is constant to $N$ and $M$
though a DMD~\cite{DMD1} contains $O(N^2T)$ parameters in $T$ layers.

The computational complexity of TPG-detector  per iteration 
is $O(MN)$ because one needs to calculate the vector-matrix products $Hs_{t}$ and $W(y-Hs_t)$
that take $O(MN)$ computational steps.
We need to calculate
the pseudo inverse matrix $W$ taking $O(M^3)$ computational steps 
only when $H$ changes.


\begin{figure}[bt]
        \begin{center}
          \includegraphics[clip, width=6.9cm]{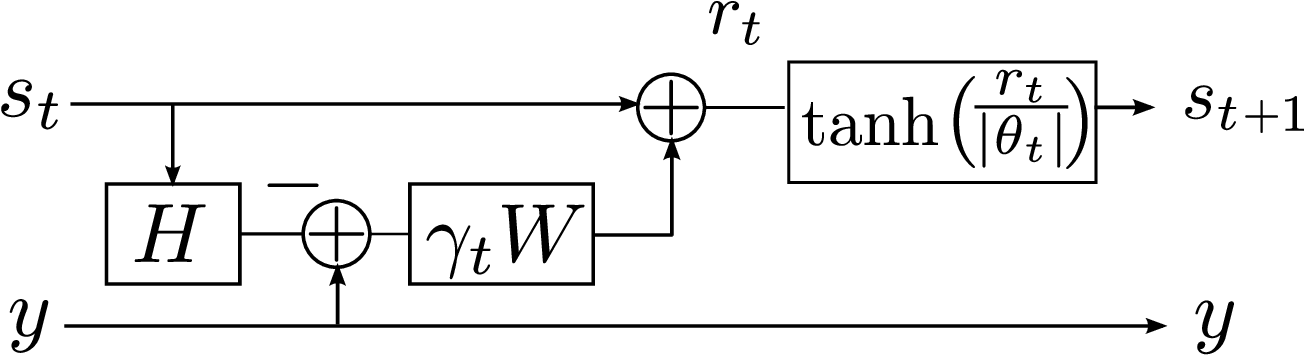}
          \caption{The $t$th layer of the TPG-detector. The trainable parameters are $\gamma_t$ and $\theta_t$.}
          \label{fig:TPG-detector}
        \end{center}
\end{figure}

The TPG-detector is trained based on the incremental training described in Section~\ref{sec_tpg}.
The training data is generated randomly according to the channel model with fixed variance $\sigma_w^2$ corresponding to a given SNR.
As described in Section~\ref{sec_set}, we assume a practical situation in which a channel matrix $H$ is a random variable.
According to this assumption, a matrix $H$ is randomly generated for each mini-batch in a training process of TPG-detector.

\section{Numerical Results}
In this section, we show the detection performance of TPG-detector and compare it to that of known 
algorithms such as IW-SOAV which is known as one of the most efficient iterative algorithms 
for massive overloaded MIMO systems.

\subsection{Experimental setup}

A transmitted vector $x$ is generated uniformly at random.
The BER is then evaluated for a given SNR.
We use randomly generated channel matrices for BER estimation. 

TPG-detector was implemented by PyTorch 0.4.0 \cite{PyTorch}.
The following numerical experiments were carried out on a PC with GPU NVIDIA GerForce GTX 1080 and Intel Core i7-6700K CPU 4.0GHz $\times$ 8.
In this paper, a training process is executed with $T=50$ rounds using the Adam optimizer~\cite{Adam}. 
A training process takes within 25 minutes under our environment.
To calculate the BER of TPG-detector, a sign function $\mathrm{sgn}(z)$ 
which takes $-1$ if $z\le 0$ and $1$ otherwise is applied to the final estimate $s_{T+1}$.

As the baselines of detection performance, we use 
the ERTS~\cite{ERTS}, IW-SOAV~\cite{IW-SOAV2}, and the standard MMSE detector.
The ERTS is a heuristic algorithm based on a tabu search for overloaded MIMO systems.
 The parameters of ERTS is based on~\cite{ERTS}.
The IW-SOAV is a double loop algorithm 
whose inner loop is the W-SOAV optimization recovering a signal using a proximal operator.
Each round of the W-SOAV takes $O(MN)$ computational steps, which is comparable to that of TPG-detector.
After finishing an execution of the inner loop with $K_{\mathrm{itr}}$ iterations,
several parameters are then updated in a re-weighting process based on a tentative recovered signal.
This procedure is repeated $L$ times in the outer loop.
The total number of steps of the IW-SOAV is thus $K_{\mathrm{itr}} L$.
In the following, we use the simulation results in~\cite{IW-SOAV2} with $K_{\mathrm{itr}}=50$.

\subsection{Main results}

\begin{figure}[tb]
  \centering
  \includegraphics[width=7.3cm]{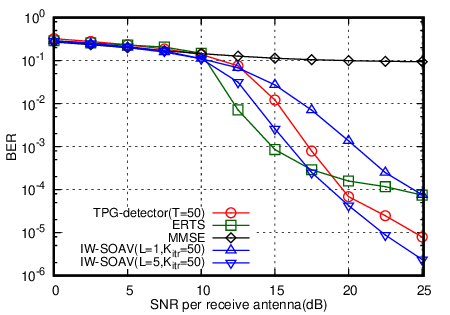}
  \caption{BER performance for $(N,M) = (200,128)$.}
  \label{fig:BER_N200}  
\end{figure}

\begin{figure}[tb]
  \centering
  \includegraphics[width=7.3cm]{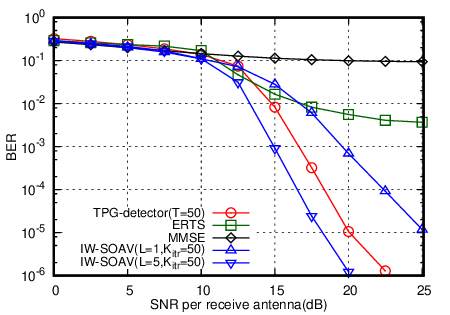}
  \caption{BER performance for $(N,M) = (300,192)$. }
  \label{fig:BER_N300}    
\end{figure}

We first present the BER performance of each detector as a function of SNR for $(N,M)=(200,128)$ 
in Fig.~\ref{fig:BER_N200}.
The results show that the MMSE detector fails to detect transmitted signals reliably (BER $\simeq 10^{-1}$) 
because the system is underdetermined.
The ERTS detector shows the best BER performance in a middle SNR region 
where SNR is between $10$ dB and $20$ dB.
On the other hand, TPG-detector exhibits the BER performance superior to that of the IW-SOAV ($L=1$), 
i.e., TPG-detector achieves approximately $5$ dB gain at $\mathrm{BER}=10^{-4}$ over the IW-SOAV ($L=1$).
Note that the computational cost for executing TPG-detector with $T=50$ is almost comparable 
to that of the IW-SOAV ($L=1$).
More interestingly, the BER performance of TPG-detector  
is fairly close to that of IW-SOAV {($L =5$)}.
For example, with $\mathrm{SNR}=20$ dB, the BER estimate of TPG-detector  
is $6.8 \times 10^{-5}$ whereas that of the IW-SOAV {($L=5$)}
is $4.3 \times 10^{-5}$.  It should be noted that the total number iterations of the IW-SOAV ($L=5$) is 250.

Figure \ref{fig:BER_N300} shows the BER performance for $(N,M)=(300,196)$.
In this case, ERTS shows relatively poor BER performance without a narrow region.
TPG-detector successfully recovers transmitted signals with lower BER than that of the IW-SOAV($L=1$).
It again achieves about $5$ dB gain against the IW-SOAV($L=1$) at $\mathrm{BER}=10^{-5}$.
Although the IW-SOAV {($L = 5$)} shows considerable performance improvements in this case,
the gaps between the curves of TPG-detector and the IW-SOAV {($L=5$)} 
 are about 2 dB at $\mathrm{BER}=10^{-5}$.

\begin{figure}[tb]
  \centering
  \includegraphics[width=7.3cm]{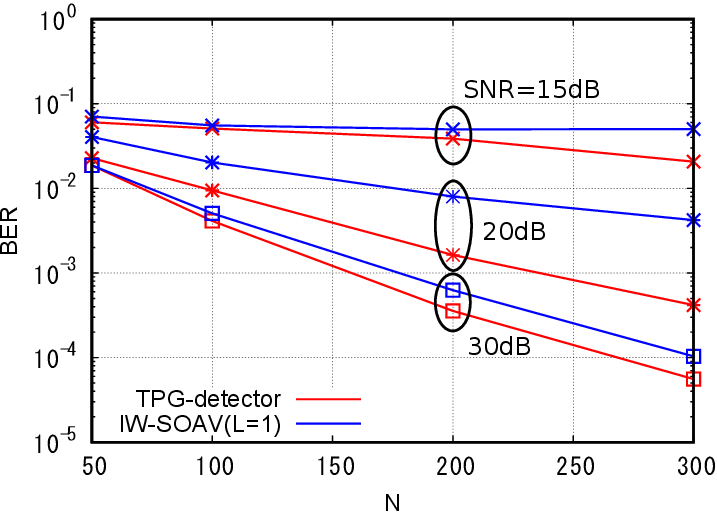}
  \caption{Antenna size dependency of BER performance for fixed rate $M/N=0.6$.}
  \label{fig:BER_N}
\end{figure}

{In Fig.~\ref{fig:BER_N}, we show the BER performance of TPG-detector and IW-SOAV ($L=1$) for some antenna sizes $N$ with the rate $M/N=0.6$ fixed.
The gap of their BER performances is especially large for SNR$=20$ (dB).
We also find that the gain of the TPG-detector increases as $N$ grows though these algorithms have the same computational costs.
It is confirmed that the TPG-detector outperforms other low-complexity algorithms especially in the massive overloaded MIMO channels.}

\begin{figure}[tb]
  \centering
  \includegraphics[width=7.3cm]{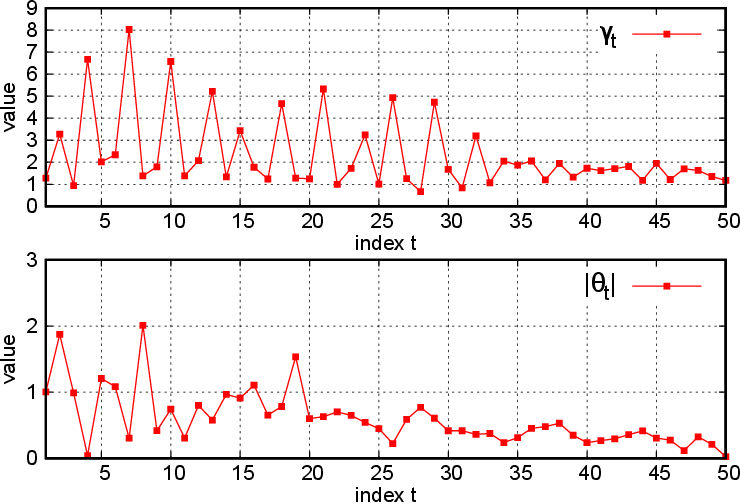}
  \caption{Sequences of learned parameters $\gamma_t$ (upper) and $|\theta_t|$ (lower); 
  $(N,M)=(300,192)$, $\mathrm{SNR}=20$ (dB), $1\le t\le T=50$.}
  \label{fig:parameter}
\end{figure}

Finally, Fig.~\ref{fig:parameter} displays the learned parameters {$\{\gamma_t,|\theta_t|\}$} of TPG-detector   
after a training process as a function of a layer index $t(=1,\dots,T)$.
We find that they exhibit a zigzag shape with damping amplitude similar to that observed in TISTA~\cite{TISTA}.
The parameter $\gamma_t$, the step size of a linear estimator, is expected to accelerate the convergence of the signal recovery.
Theoretical treatments for providing reasonable interpretation on 
these characteristic shapes of the learned parameters 
are interesting open problems.

\section{Concluding remarks}

In this paper, we proposed TPG-detector, a deep learning-aided iterative decoder for massive overloaded MIMO channels.
TPG-detector contains two trainable parameters for each layer: $\gamma_t$ controlling a step size of the linear estimator and
 $\theta_t$ dominating strength of the nonlinear estimator.
The total number of the trainable parameters in $T$ layers 
is thus $2T$, which is significantly smaller than that used in the previous studies such as \cite{DMD1,DMD2}.
This fact promotes fast and stable training processes for TPG-detector.
The numerical simulations show that TPG-detector outperforms the state-of-the-art IW-SOAV ($L=1$) by a large margin and
achieves a comparable detection performance to the IW-SOAV {($L=5$)}.
TPG-detector therefore can be seen as a promising iterative detector for overloaded MIMO channels 
providing an excellent balance between a low computational cost and a reasonable detection performance.
There are several open problems regarding this study.
First, adding a re-weighting process similar to the one used in the IW-SOAV ($L\ge 2$)
to TPG-detector seems an interesting direction to improve the detection performance.
Secondly, enhancing TPG-detector toward a large constellation such as QAM is a practically important problem.

\section*{Acknowledgement}
The authors are very grateful to Mr.~Ryo Hayakawa 
for {providing numerical data and simulation programs in~\cite{IW-SOAV2}.}
This work was partly supported by JSPS Grant-in-Aid for Scientific Research (B) 
Grant Number 16H02878 (TW) and Grant-in-Aid for Young Scientists (Start-up) Grant Number 17H06758 (ST).



\end{document}